\documentclass[12pt]{article}

\usepackage{times}
\usepackage{fixltx2e}
\usepackage{graphicx}
\usepackage[table]{xcolor}
\usepackage{ragged2e}
\usepackage{multirow}
\usepackage{helvet}
\usepackage{booktabs}
\usepackage{colortbl}
\RequirePackage{mathpazo}
\usepackage{amsmath}
\usepackage[svgnames]{xcolor}

\usepackage{blindtext}

\usepackage[colorlinks = true,
            linkcolor = magenta,
            urlcolor  = black,
            citecolor = magenta,
            anchorcolor = magenta]{hyperref}

\usepackage{xurl}

\makeatletter
\newenvironment{figurehere}
{\def\@captype{figure}}
{}
\makeatletter

\usepackage{ccaption}
\usepackage{siunitx}
\usepackage{fixltx2e}
\newenvironment{sciabstract}{%
\begin{quote} \bf}
{\end{quote}}

\newcommand\std[1]{\textcolor{black}{#1}}
\newcommand\stdfig[1]{\textcolor{darkgray}{#1}}

\newcommand\yu[1]{\textcolor{Indigo}{#1}}
\usepackage{bm}

\newcounter{lastnote}

\title{BEOL Ferroelectric Compute-in-Memory Ising Machine for Simulated Bifurcation}

\author
{
Yu Qian$^{1*}$, Alptekin Vardar$^{2*}$, Konrad Seidel$^{2}$, David Lehninger$^{2}$\\ 
Maximilian Lederer$^{2}$, Zhiguo Shi$^{1}$, Cheng Zhuo$^{1}$, \\ Kai Ni$^{3\dagger}$, Thomas K{\"a}mpfe$^{2,4\dagger}$, Xunzhao Yin$^{1\dagger}$
\\
\normalsize{$^{1}$Zhejiang University, Hangzhou, China;} \\
\normalsize{$^{2}$Fraunhofer IPMS, Dresden, Germany;}\\
\normalsize{$^{3}$University of Notre Dame, Notre Dame, USA;}\\
\normalsize{$^{4}$TU Braunschweig, Germany;}
\\
\normalsize{$^\ast$Equal contributions;} \\
\normalsize{$^\dagger$To whom correspondence should be addressed; E-mail:}\\ 
\footnotesize{czhuo@zju.edu.cn, kni@nd.edu,  thomas.kaempfe@ipms.fraunhofer.de, xzyin1@zju.edu.cn.}
}

\date{}

\begin{document} 

\maketitle 
\section*{Abstract}
\begin{sciabstract}
\std{
Computationally hard combinatorial optimization problems are pervasive in science and engineering, yet their NP-hard nature renders them increasingly inefficient to solve on conventional von Neumann architectures as problem size grows. Ising machines implemented using dynamical, digital and compute-in-memory (CiM) approaches offer a promising alternative, but often suffer from poor initialization and a fundamental trade-off between algorithmic performance and hardware efficiency. Hardware-friendly schemes such as simulated annealing converge slowly, whereas faster algorithms, including simulated bifurcation, are difficult to implement efficiently in CiM hardware, limiting both convergence speed and solution quality. 
To address these limitations, Here we present a ferroelectric field-effect transistor (FeFET)–based CiM Ising framework that tightly co-designs algorithms and hardware to efficiently solve large-scale combinatorial optimization problems. The proposed approach employs a two-step algorithmic flow: an attention-inspired initialization that exploits global spin topology and reduces the required iterations by up to 80\%, followed by a lightweight simulated bifurcation algorithm specifically tailored for CiM implementation. To natively accelerate the core vector–matrix and vector–matrix–vector operations in both steps, we fabricate a 32$\times$256 FeFET CiM chip using ferroelectric capacitors integrated at the back end of line of a 180-nm CMOS platform. Across Max-Cut instances with up to 100,000 nodes, the proposed hardware–software co-designed solver achieves up to a 175.9$\times$ speedup over a GPU-based simulated bifurcation implementation while consistently delivering superior solution quality. 
}

\end{sciabstract}

\section*{Introduction}
\label{sec:introduction}

\vspace{-3ex}
\std{Combinatorial optimization problems (COPs) pervade real-world decisions in 
domains such as logistics, resource distribution, communication network design, finance, drug identification, and transportation systems, among others \cite{yu2013industrial, paschos2014applications, naseri2020application, barahona1988application}.}
\std{Many of these problems are non-deterministic polynomial-time-hard (NP-hard) problems, representing some of the most 
challenging tasks within the NP class. 
When tackled on conventional Von-Neumann digital computers, the computational cost and latency of COP solvers typically grow exponentially with problem size
\cite{markov2014limits, markov2013know, greenlaw1995limits}, leading to a widening gap between practical solving demand and available compute.}
\std{These limitations motivate the exploration of 
alternative architectures and algorithms that can solve large-scale and complex COPs with substantially improved computational efficiency.
}

\std{A  broad class of COPs, including graph coloring, Max-Cut, and traveling salesman problem, etc., can be formulated as Ising model \cite{lucas2014ising}, where problem variables are represented as binary spins and constraints are captured through pairwise spin couplings. 
The corresponding objective function of the problem is expressed as the Ising Hamiltonian energy.
Solving the problem involves finding the spin configuration that minimizes the corresponding Ising Hamiltonian $H_{\mathrm{P}}$:}
\begin{equation}
\vspace{-1ex}
\label{equ:Ising model1}
    \min H_{\mathrm{P}}=\sum_{i,j=1}^{N}J_{i j}\sigma_{i}\sigma_{j} + \sum_{i=1}^{N}h_{i}\sigma_{i}
\end{equation}
\std{where $N$ denotes the number of spins,  $\sigma_i \in\{1, -1\}$ represents the state of spin $i$. $J_{ij}$ and $h_i$ stand for the coupling between spin $i$ and $j$ and the self-coupling of spin $i$, respectively. 
Designing hardware that can efficiently navigate the energy landscape of this objective is therefore a promising route toward scalable COP solvers.}

\std{
To this end,  customized hardware-algorithm co-designed frameworks, known as  Ising machines, have been proposed to approximately minimize the
Ising Hamiltonian and thus correspondingly find the optimal solution to the original COPs, as shown in Fig. \ref{fig:motivation}\textbf{a}. 
Existing Ising machines can be broadly classified into three categories as shown in Fig. \ref{fig:motivation}\textbf{b} and \textbf{c}: dynamical system-based, compute-in-memory (CiM) system-based and digital system-based implementations. 
The dynamical system-based Ising machines exploit the intrinsic physical dynamics and natural tendency of nonlinear systems
to relax toward low-energy states
\cite{cilasun2025coupled, moy20221, ahmed2021probabilistic,dutta2021ising, roychowdhury2021bistable, mallick2023cmos, afoakwa2021brim, pierangeli2019large, honjo2021100, yamamoto2020coherent, mcmahon2016fully, bohm2019poor}. 
While they can be highly energy efficient, they often suffer from limited scalability and weak robustness, as  slight deviations in the  implementations of spin couplings can disrupt convergence
\cite{ahmed2021probabilistic, mallick2023cmos}. 
As these solvers rely on intrinsic physical dynamics for convergence, 
they operate without computing units for explicit arithmetic computation, 
thereby rendering them inherently incapable of deploying other algorithms. 
Digital Ising machines, by contrast, offer strong robustness and programmability, and can support a wide spectrum of algorithms \cite{yamamoto20207, katsuki2022fast, onizawa2023local,takemoto20214, tatsumura2021scaling}, but their throughput and energy efficiency are fundamentally limited by data movement between memory and compute. 
CiM  systems-based Ising machines alleviate this bottleneck by performing vector-matrix operations directly in memory arrays  \cite{yue2024scalable, yin2024ferroelectric, qian2025ferroelectric, qian2024c, qian2024hycim, qian2025device, dee2024design}, enabling massively parallel evaluation of the Ising energy.
On the algorithmic side, most existing digital and CiM Ising machines 
implement heuristic algorithms such as 
simulated annealing (SA) \cite{yue2024scalable, jiang2023efficient, yamamoto20207, yin2024ferroelectric, qian2025ferroelectric} and simulated bifurcation (SB) \cite{goto2019combinatorial, goto2021high, zhang2024high, dee2024design}. 
SA is naturally compatible with binary spin representations and thus straightforward to implement in hardware, but typically requires a large number of iterations to reach high-quality solutions
\cite{yamamoto20207,katsuki2022fast, onizawa2023local, takemoto20214, jiang2023efficient, yue2024scalable, yin2024ferroelectric,qian2024hycim,qian2025ferroelectric,qian2024c,qian2025device}.
SB, in contrast, has been shown to converge faster and achieve superior solution quality, especially for large-scale COPs, for example, reporting over 
100$\times$ speedup improvement in solving 2000-node Max-Cut problems \cite{goto2019combinatorial, goto2021high, zhang2024high, dee2024design},
making it a highly attractive 
algorithmic basis for  Ising machines. 
}
Despite its huge potential, most current SB implementations are predominantly confined to digital systems  \cite{goto2019combinatorial, goto2021high, zhang2024high}, which suffer from the inherent latency and energy penalties of the Von Neumann memory bottleneck. 
To overcome this, mapping SB onto CiM architectures is essential for unlocking its full efficiency. 
Yet, existing efforts in this direction have been limited to simulation-level validations \cite{dee2024design}, and the feasibility of a physical hardware accelerator has not yet been experimentally proven.

\begin{figurehere}
	\centering
	\includegraphics [width=1\linewidth]{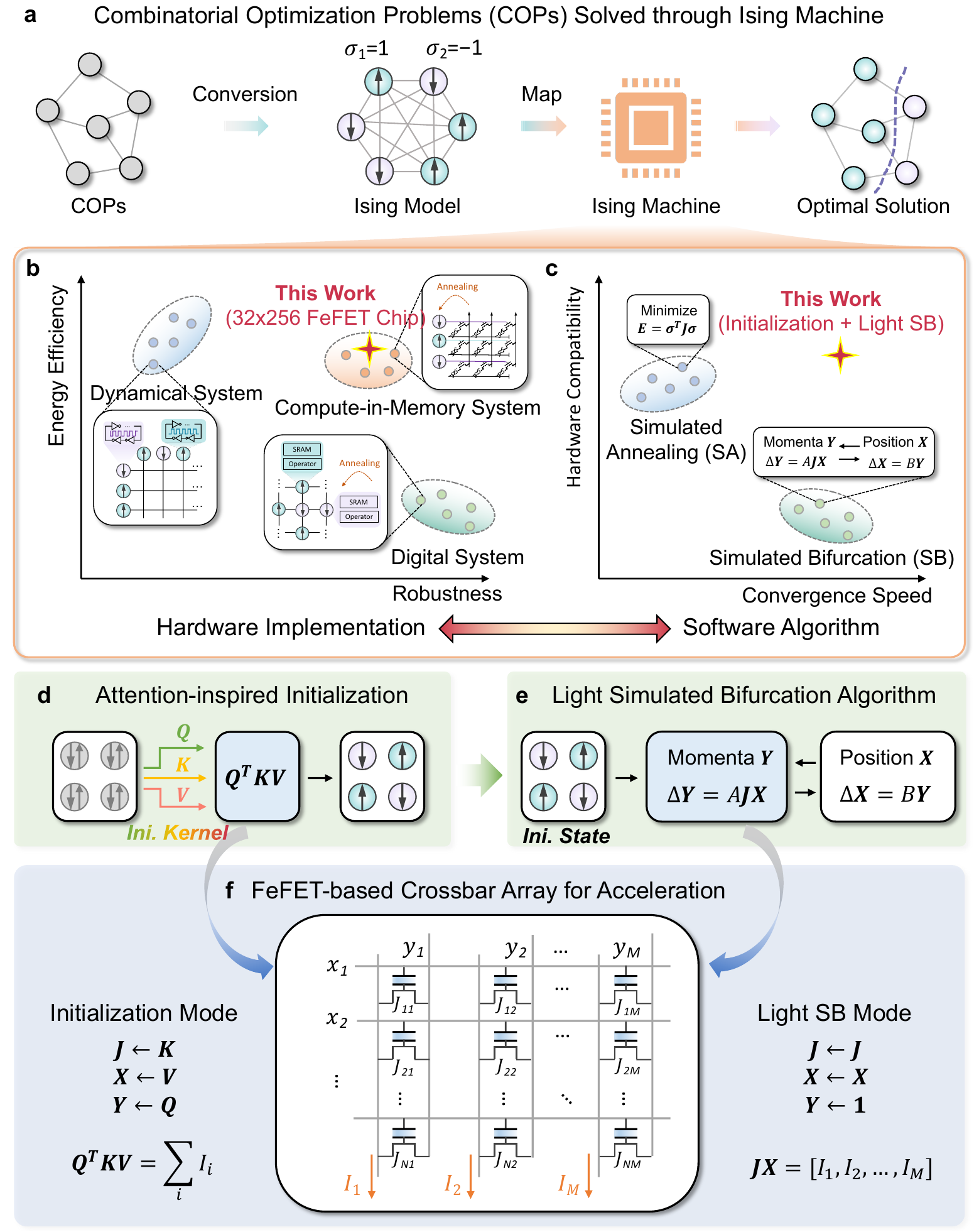}
	\caption{
    \stdfig{
    \textbf{Acceleration of COP solving through hardware-software co-design.} 
    \textbf{a.} COPs can be converted to an Ising model, and then mapped to an Ising machine for optimal solution finding. 
    \textbf{b.} Existing Ising machines face challenges in hardware implementation, exhibiting either low energy efficiency or weak robustness. 
    \textbf{c.} Concurrently, current software algorithms are often not hardware-friendly or converge slowly. 
    These limitations motivate our proposed work, which involves \textbf{d.} an attention-inspired initialization algorithm and \textbf{e.} a light simulated bifurcation algorithm for rapid and improved solution discovery with lower hardware overhead. 
    \textbf{f.} The core operations within the algorithm flow are further accelerated by our  fabricated FeFET-based CiM crossbar prototype to improve solving efficiency. 
    }}
	\label{fig:motivation}
\end{figurehere}

\std{
While both  digital and CiM solvers support various algorithms and offer hardware advantages, these existing Ising machines still face two critical challenges in hardware-software co-design. 
First, there is an algorithmic deficiency in initialization: 
Almost all existing works rely on 
random spin initialization for COPs \cite{yin2024ferroelectric, qian2025ferroelectric, goto2019combinatorial, goto2021high},
disregarding the underlying graph topology.
Such topology-agnostic initialization can severely slow convergence  and degrade solution quality, 
leading to suboptimal solving efficiency \cite{hassanat2018improved}.
Second, 
the standard SB formulation is poorly aligned with hardware constraints.
Although SB offers substantial advantages over SA, it is hardware-unfriendly:
It doubles the number of variables by representing each original variable with both position and momentum components, 
requires floating-point precision compared to the original binary variables, 
and involves computationally expensive cubic non-linear terms that exceed the complexity of standard quadratic computations 
(Section. \ref{subsec: SB} in Supplementary) \cite{goto2019combinatorial, goto2021high}.
These factors severely complicate efficient hardware realization and limit the attainable energy and area efficiency.
}

\std{
In this article, we report a CiM-based Ising machine that addresses both limitations through a tightly integrated hardware-algorithm co-design. 
We propose a two-step algorithmic flow that jointly improves convergence speed,  hardware compatibility and  solution quality.
The first step is an attention-inspired initialization scheme that leverages the global spin topology to construct informative initial spin states for COPs.
This scheme is generic and can be seamlessly integrated in both SA and SB,  reducing the number of required iterations by 80\% and  significantly accelerating the convergence, as shown in Fig. \ref{fig:motivation}\textbf{d}. 
The second step is a hardware-friendly light SB algorithm as illustrated in Fig. \ref{fig:motivation}\textbf{e}, which removes hardware-intensive terms and adopts ternary quantization, thereby simplifying computational operations and reducing memory overheads while preserving solution quality when combined with the new initialization.
}

\std{
To support these algorithmic advances, 
we fabricate the first $32\times256$ back-end-of-line (BEOL) ferroelectric CiM chip dedicated to Ising computation   
(Fig. \ref{fig:motivation}\textbf{f}).
The FeFET-based CiM array accelerates the key  vector-matrix multiplication (VMM) and vector-matrix-vector multiplication (VMV) operations required by both the attention-inspired initialization and the light SB algorithms. 
When evaluated on 
large-scale Max-Cut problems with up to 100,000 nodes, the proposed Ising machine achieves up to 175.9$\times$ speedup to convergence 
and even modest
improvements in solution quality 
compared with a GPU-based implementation of conventional SB. 
}

\vspace{-2ex}
\section*{FeFET based CiM architecture}
\label{sec:hardware}

\std{
Recognizing the promise of the FeFET crossbar in accelerating both VMM and VMV multiplication operation \cite{yin2024ferroelectric}, we fabricate a 32$\times$256 FeFET CiM array prototype, as shown in Fig. \ref{fig:chip}. 
The  chip is integrated onto a 180nm technology platform. The formation of a FeFET leverages a front end Si transistor and a back end of line (BEOL) metal-ferroelectric-metal (MFM) capacitor, making the device an effective metal-ferroelectric-metal-insulator-semiconductor (MFMIS) FeFET. It is well known that such a device allows the optimization of the device performance through independent tuning of the MFM capacitor area with respect to the transistor area \cite{ni2018soc}. 
Fig. \ref{fig:chip}\textbf{b} shows the 3D schematic of the FeFET device including metal layers from M1 to M5, with  detailed transmission electron microscopy (TEM) images from two vertical cross-sections  in Fig. \ref{fig:chip}\textbf{a} and \textbf{b}, respectively. 
In our design, the MFM capacitor is integrated on M4, as shown in the zoom-in TEM image. 
Elemental analysis further confirms the presence of Hafnium (Hf) in the ferroelectric layer and Titanium (Ti) in the electrodes, as shown in Fig. \ref{fig:chip}\textbf{d}. 
Based on this device,  a 32$\times$256 FeFET CiM array layout is designed as shown in Fig. \ref{fig:chip}\textbf{e}. 
The fabricated chip micrograph, including the core array and read/write peripherals, is presented in Fig. \ref{fig:chip}\textbf{f}. }

\begin{figurehere}
	\centering
	\includegraphics [width=1\linewidth]{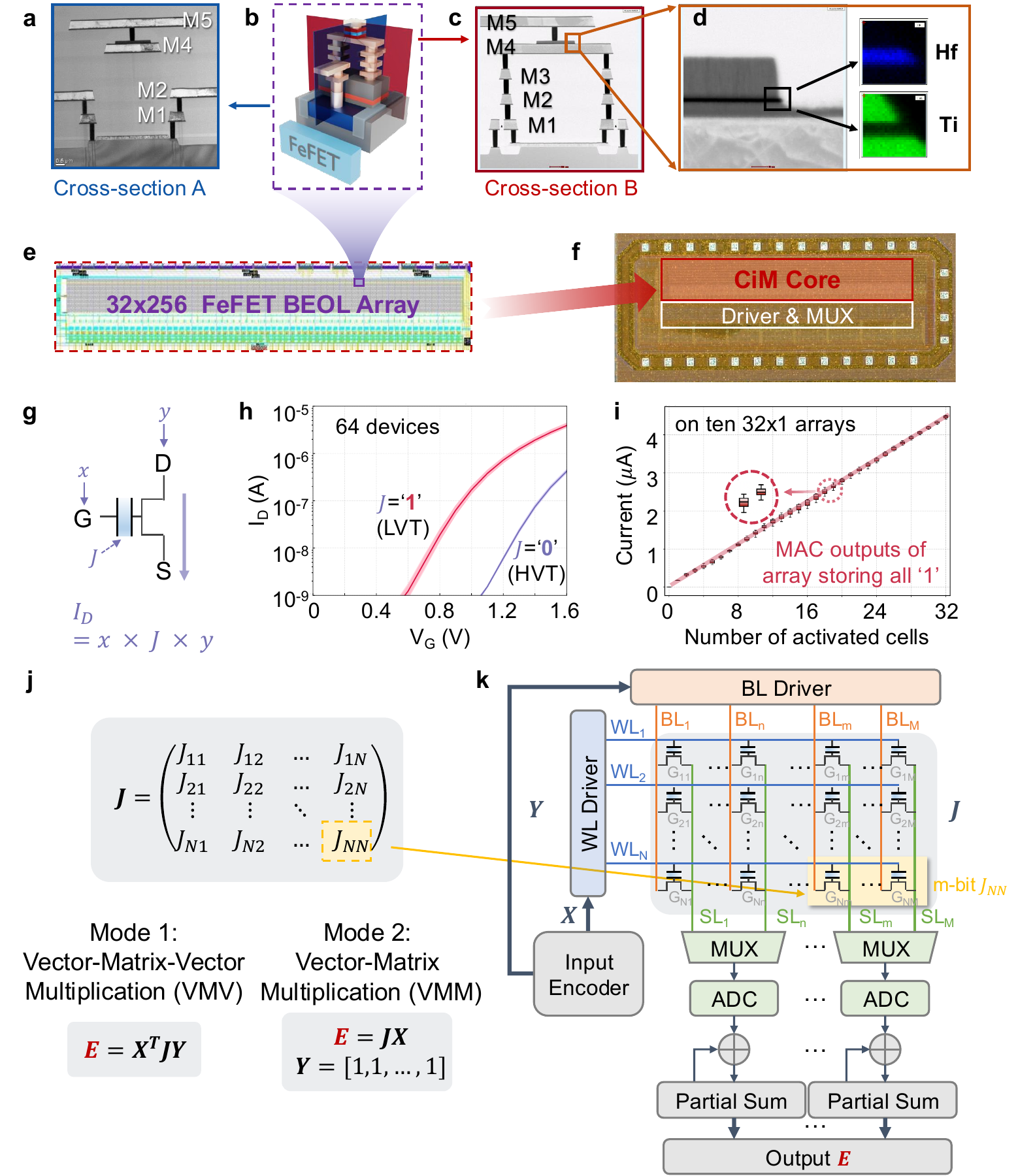}
 \caption{ 
 \stdfig{
 \textbf{FeFET-based CiM array for the acceleration of vector-matrix-vector (VMV) and vector-matrix multiplication (VMM) operation.}
 \textbf{a.} 3D schematic of a single FeFET device with transmission electron microscopy (TEM) images from two vertical cross-sections. 
 \textbf{b.} Layout of a 32$\times$256 FeFET BEOL array. 
 \textbf{c.} Micro-graph of the fabricated chip consisting of core array and read/write peripherals. 
 \textbf{d.} Single FeFET supports the function of three scalar product $x\times J \times y$.
 \textbf{e.} Characteristics of 60 FeFETs for the two memory states.
 \textbf{f.} Measured column currents from 10 arrays show a good linearity with respect to the number of activated cells in the column, thus promising for VMV and VMM operations.
 \textbf{g.} Both VMV and VMM operations can be mapped onto \textbf{h.} FeFET-based CiM array architecture.
 }
 }
	\label{fig:chip}
\end{figurehere}

\std{
The intended VMV and VMM operations all rely on the fact that a single FeFET inherently realizes the function of scalar product $x\times J \times y$, as illustrated in \ref{fig:chip}\textbf{g}. 
Here data $J$ is stored as the threshold voltage (\textit{V}\textsubscript{TH}) state within the device, 
while inputs $x$ and $y$ are applied to the gate (G) and drain (D), respectively.
The resultant output current flowing from drain to source, i.e., $I_{D}$, directly represents the scalar product, i.e., $I_{D}=x\times J \times y$. 
To program \textit{V}\textsubscript{TH} of a FeFET,  +4V/-4V gate pulses with 1$\mu$s duration are applied to encode $J$=1/0 into the device, respectively, which orient the ferroelectric polarization  toward the channel/gate-metal, and hence sets the \textit{V}\textsubscript{TH} of FeFET into a low-\textit{V}\textsubscript{TH} (LVT) state (i.e., $J$ = 1)/high-\textit{V}\textsubscript{TH} (HVT) state (i.e., $J$ = 0). 
Fig. \ref{fig:chip}\textbf{h} shows the experimentally measured 
$I_D$-$V_G$ characteristics for the two polarization states across 64 nominally identical devices. 
For an appropriate read gate bias, i.e., input $x$, the resultant current $I_D$ realizes the desired scalar product. 
Leveraging this single-device characteristic, each column in the fabricated FeFET array naturally performs a multiplication and accumulation (MAC) operation. 
Fig. \ref{fig:chip}\textbf{i} demonstrates that the column current  the array increases linearly  with the number of activated FeFETs across 10 distinct arrays,  validating the intended MAC functionality of the design. 
}

\std{
We then present the implementation of both VMV and VMM multiplication operations. 
As shown in Fig. \ref{fig:chip}\textbf{j}, the matrix $\bm{J}$ has a size of $N\times N$, and 
the inputs $\bm{X}$ and $\bm{Y}$ are $1\times N$ vectors whose element takes the value  1/-1 in VMM and 1/0 in VMV.
The object function for VMV multiplication operation is $E=\bm{X^TJY}$. 
For VMM operation, $E=\bm{JX}$ is computed by setting $\bm{Y}$ to a unity vector, i.e., $\bm{Y}=[1,...,1]$. 
For hardware implementation, the matrix $\bm{J}$ is mapped onto the crossbar array with m-bit quantization, as illustrated in Fig. \ref{fig:chip}\textbf{k}. 
Each FeFET cell  stores one bit of the matrix element, so an $N\times N$ matrix $\bm{J}$ requires an $N\times (N\times m)$ crossbar array. 
Cells within each column share a bit line (BL) and source line (SL), while cells within each row share a word line (WL). 
The input vectors $\bm{X}$ and $\bm{Y}$ are mapped to the WL and SL, respectively. 
The resulting column output currents are directed to multiplexers (MUXs) and  analog-to-digital converters (ADCs), then processed by shift-and-add units to obtain the final output $E$. 
}

\section*{Attention-inspired Initialization and Light SB}
\label{sec:software}

\std{
The initialization process is a critical step in COP solving,
However, most existing approaches  rely on random initialization \cite{yin2024ferroelectric, qian2025ferroelectric, goto2019combinatorial, goto2021high}, where spins are randomly assigned $\pm1$. 
Such a strategy ignores the underlying interaction topology among variables, and consequently increases the number of iterations required for subsequent algorithms to converge. 
This limitation motivates a solution inspired from the self-attention mechanism \cite{vaswani2017attention}.
In self-attention, 
the importance of each input element (e.g., the $n^{th}$ token in a sentence) 
is evaluated  not only with respect to its immediate  neighbors (i.e., the $(n-1)^{th}$ and $(n+1)^{th}$ tokens), but also with respect to all other tokens in the sentence. 
In other words, 
even indirectly related variables contribute to the computation when evaluating the score of a given input element. 
Inspired by this principle, we propose a universal initialization scheme for COPs that explicitly considers the global spin topology. 
}

\std{
Fig. \ref{fig: algorithm flow}\textbf{a} illustrates the operation flow of attention-inspired initialization scheme. 
For a given Ising model represented by an $N\times N$ matrix $\bm{J}$, we first construct three $N\times N$ matrices, $\bm{K}$, $\bm{Q}$ and $\bm{V}$.
The matrix $\bm{K}$ directly captures the overall connectivity  of the COP and is set equal to $\bm{J}$.
In contrast, the matrix $\bm{Q}$ encodes the disconnection information.
Specifically, each column vector $\bm{Q_i}$ of matrix $\bm{Q}$ is defined such that its elements $\bm{Q_i}[j]$ is $1$ if spins $i$ and $j$ are not connected (i.e., $\bm{K}[j,i]==0$), and $0$ otherwise. 
This means that $\bm{Q}$ is the logical complement of $\bm{K}$'s connection pattern. 
The matrix $\bm{V}$ is also composed of column vectors, with each column vector $\bm{V_i}=\bm{K[:,i]}$, capturing the connection neighbors of spin $i$. 
Based on $\bm{K}$, $\bm{Q}$ and $\bm{V}$, we define a score $\bm{S}_i$ for each spin $i$: 
\begin{equation}
\label{equ: score}
    \bm{S}_i = \bm{Q_i}^T \bm{K} \bm{V_i}=\sum_{j,k}\bm{Q_i}[j]\bm{K}[j,k]\bm{V_i}[k]
\end{equation}
}
\std{
\noindent which can be efficiently accelerated on the FeFET-based CiM chip by storing $\bm{K}$ in the crossbar and applying $\bm{Q_i}^T$ and $\bm{V_i}$ along BL and SL, respectively,  as described in the previous section. 
The value of $\bm{S}_i$ directly correlates with the non-zero element overlaps across $\bm{Q_i}$, $\bm{K}$, and $\bm{V_i}$. 
A positive contribution to $\bm{S}_i$ only occurs when
$\bm{Q_i}[j]$, $\bm{K}[j,k]$ and $\bm{V_i}[k]$ are all $1$. 
This corresponds to the case where spin $j$ is a common neighbor of both spin $i$ and $k$, while spin $i$ and $k$ themselves are disconnected. 
Hence, a higher score $\bm{S}_i$ indicates that spin $i$ has  a higher ratio of its second-order neighbors (neighbors-of-neighbors) relative to its first-order neighbors. 
This observation informs our initialization strategy: 
we group high-score spins (i.e., local spins and their second-order neighbors) to maximize the connections (i.e., edge numbers) between them and their common first-order neighbors. 
The  initial state $\sigma_i$ of spin $i$ is then assigned by comparing its score $\bm{S}_i$ with the average score $Mean(\bm{S})$: 
\begin{equation}
\label{equ: initial state}
    \sigma_i = \begin{cases}
	1,\bm{S}_i \geq Mean(\bm{S})\\
	0,\bm{S}_i < Mean(\bm{S})
	\end{cases}
\end{equation}
}

\begin{figurehere}
	\centering
	\includegraphics [width=1.0\linewidth]{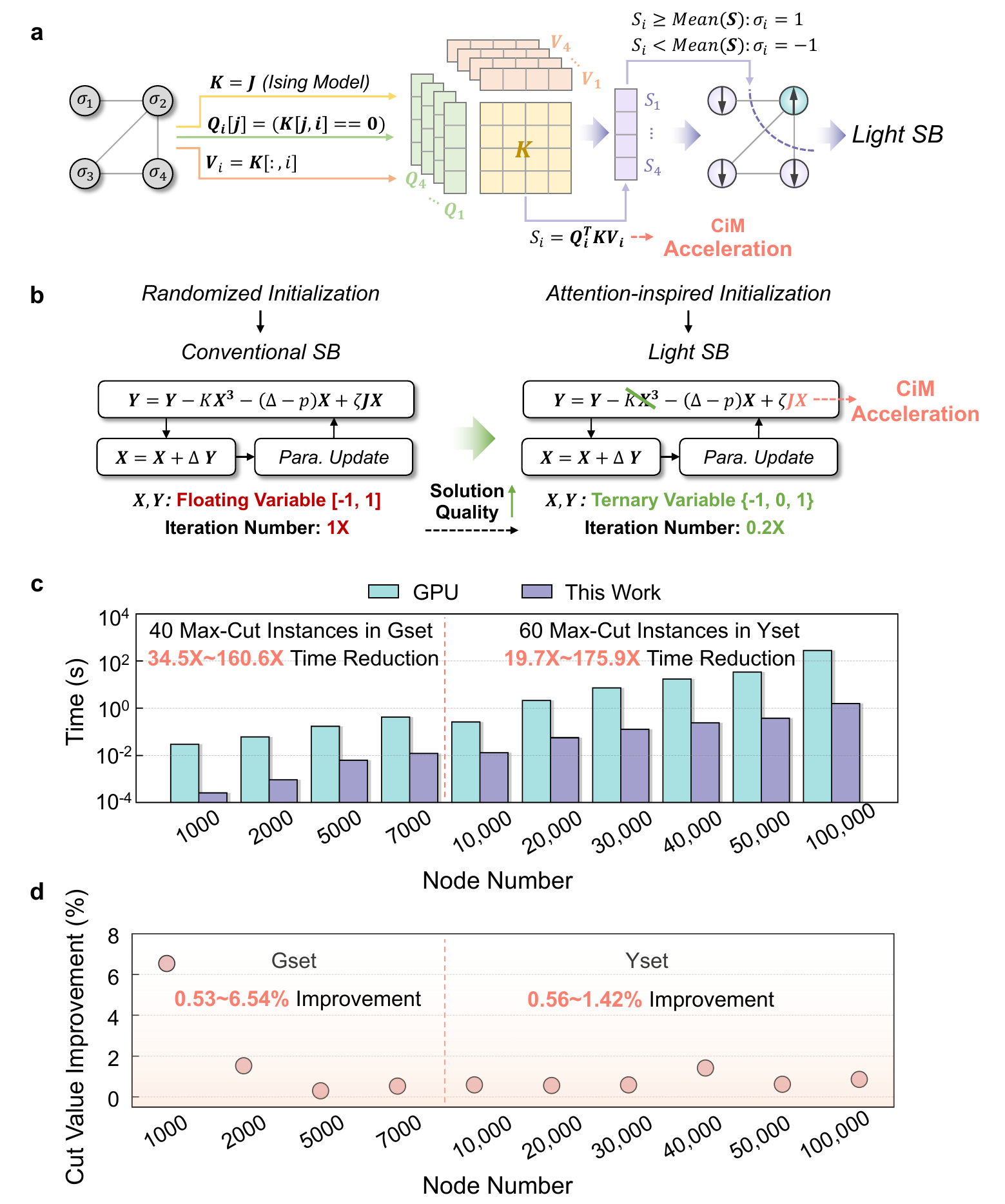}
	\caption{ 
    \stdfig{
    \textbf{Working flow of the proposed Ising machine.} 
    \textbf{a.} In the attention-inspired initialization process, three matrices, i.e., $\bm{K}$, $\bm{Q}$ and $\bm{V}$, are generated based on spin connection topology given a COP. 
    Each spin's score, $S_i = \bm{Q_i^T} \bm{K} \bm{V_i}$, is calculated with CiM acceleration. 
    The initial state of spin $i$ is then determined by comparing its score $S_i$ to the average of all scores $Mean(\bm{S})$. 
    \textbf{b.} Following initialization, the light SB algorithm is executed. Compared to conventional SB, light SB reduces input quantization from floating-point to ternary, eliminates the computationally expensive cubic term $\bm{X}^3$, and accelerates the core VMM multiplication operations within the CiM architecture. 
    This leads to an 80\% reduction in iterations compared to conventional SB. 
    \textbf{c.} In solving 100 Max-Cut instances with 1,000 to 100,000 nodes \cite{maxcutstanford, yset},  the proposed Ising machine achieves up to 175.9$\times$ speedup to convergence and \textbf{d.} improved quality of solution compared to an A6000 GPU. 
    }
    }
    \label{fig: algorithm flow}
\end{figurehere}

\std{
Following the attention-inspired initialization, we propose a light SB algorithm tailored for efficient hardware implementation. 
In the conventional SB algorithm \cite{goto2019combinatorial, goto2021high}, each spin $i$ is represented by two continuous new variables, i.e., position $\bm{X}_i \in [-1,1]$ and momenta $\bm{Y}_i\in [-1,1]$. 
After random initialization, SB evolves by 
iteratively updating the position based on its momentum, and vice-versa: 
\begin{equation}
\label{equ: SB Y}
    \bm{Y} = \bm{Y} - K\bm{X}^3 - (\Delta - p)\bm{X} + \zeta \bm{J}\bm{X}
\end{equation}
\begin{equation}
\label{equ: SB X}
    \bm{X} = \bm{X} + \Delta \bm{Y}
\end{equation}
where $K$, $\Delta$ and $\zeta$ are constants, and $p$ is an annealing-like parameter that increases over iterations.
Additional details are provided in Section \ref{subsec: SB} of Supplementary. 
Eventually, each position variable $X_i$ converges to either $+1$ or $-1$, indicating the final value  $\sigma_i$ of spin $i$. 
However, this formulation is challenging for hardware implementation primarily due to (i)
the floating-point precision required in the original formalism for both $\bm{X}$ and $\bm{Y}$, and (ii) the computationally expensive cubic term $K\bm{X}^3$. 
To address these challenges, we propose a light SB algorithm designed to simplify computation and enhance  compatibility with CiM hardware. 
}

\std{
As Fig. \ref{fig: algorithm flow}\textbf{b} shows, 
the light SB algorithm significantly reduces hardware cost through two key modifications: (i) ternary quantization of both $\bm{X}$ and $\bm{Y}$ (i.e., $\bm{X}, \bm{Y} \in \{-1, 0, 1\}$) and  (ii) removal of the hardware-intensive 
$K\bm{X}^3$ computation term. 
To accelerate the VMM operation $\bm{JX}$ with ternarised $\bm{X}$, the computation is decomposed into two physical phases: 
first, the sub-vector corresponding to the positive inputs ($\bm{X}=+1$) is multiplied with $\bm{J}$, followed by the sub-vector for the negative inputs ($\bm{X}=-1$, in this case, actually positive inputs are applied). 
The final result is obtained by subtracting the latter partial sum from the former. 
Such approximations cause a minor loss less than 5\% in solution quality when the light SB is used alone (see Section of \ref{subsec: interval} in Supplementary). 
Combining it with our attention-inspired initialization scheme yields superior final solution quality compared to the conventional SB, as discussed later.  
}

\std{
Leveraging the hardware-algorithm co-design, we benchmarked of the proposed FeFET-based Ising machine against a conventional SB solver implemented on an A6000 GPU across 100 Max-Cut problem instances. 
The reported performance results are derived from a hardware-calibrated system-level evaluation. Specifically, the FeFET CiM core is modeled as a scalable single-chip crossbar array, with latency and energy parameters directly extracted from experimental measurements of our fabricated prototype. 
The remaining digital peripheral logic is implemented and synthesized using the TSMC 180 nm PDK to ensure rigorous power and area estimation. 
The benchmark set includes 40 Gset instances 
\cite{maxcutstanford} (1000 to 7000 nodes), and 60 larger-scale instances from our generated larger-scale Yset dataset  \cite{yset} (10,000 to 100,000 nodes). The generation code and dataset are publicly available. 
Fig. \ref{fig: algorithm flow}\textbf{c} reports the time-to-solution of both solvers. 
While both solvers show increased solution times for larger problems, our Ising machine consistently converges  faster, 
achieving  34.5-160.6$\times$ speedup on Gset and 19.7-175.9$\times$ speedup on Yset compared to the GPU.  
This substantial acceleration arises from not only the 80\% iteration reduction enabled by our attention-inspired initialization scheme, but also the superior  hardware performance of the CiM architecture. 
Furthermore, Fig. \ref{fig: algorithm flow}\textbf{d} shows that the proposed work yields improved solution quality with a 0.53-6.54\%  and a 0.56-1.42\%
increase in Max-Cut value on Gset and Yset compared to the GPU implementation, respectively. 
These enhancements are attributed to the improved initial states provided by the attention-inspired initialization scheme, 
which provides a superior initial state, 
which steers the solver dynamics toward  better optimal solution. 
}

\section*{COP Solving based on Hardware-Software Co-Design}
\label{sec:evaluation}


\std{
To demonstrate the capability of the developed  FeFET CiM Ising machine in solving COPs, we first evaluate a toy  Max-Cut example, as shown in Fig.\ref{fig:demo}\textbf{a}.
The problem graph consists of 32 nodes with an edge density of 10\%, 
and the  edge weights are integers randomly assigned from $[-4, 4]$. 
The corresponding 32$\times$32 Ising matrix,
is thermometer-coded and expanded into a 32$\times$256 matrix $\bm{J}$ for mapping onto the array, as shown in Fig.\ref{fig:demo}\textbf{b}. 
The measured current matrix, shown in Fig.\ref{fig:demo}\textbf{c}, reveals the cell current $I_{D}$ of the FeFET chip. 
Compared with the ideal matrix, this measured matrix shows a slight deviation, primarily due to non-ideal factors such as device variation and read/write disturbance.
}

\std{
The problem is mapped onto five independent FeFET chips for solving. 
The evolution of the input spin variable $\bm{\sigma}$ of the light SB algorithm is depicted in Fig. \ref{fig:demo}\textbf{d}, where the spins  converge  to an optimal solution within only 20 iterations. 
Fig.\ref{fig:demo}\textbf{e} further shows the evolution of the Max-Cut value as the proposed Ising machine solves the problem. 
Experiments conducted across the five  FeFET CiM chips demonstrate that all runs achieve better solutions than conventional SB, with 4.3 to 8.7\% improvement in Max-Cut value, while converging within 900 ns. 
}

\begin{figurehere}
	\centering
	\includegraphics [width=1\linewidth]{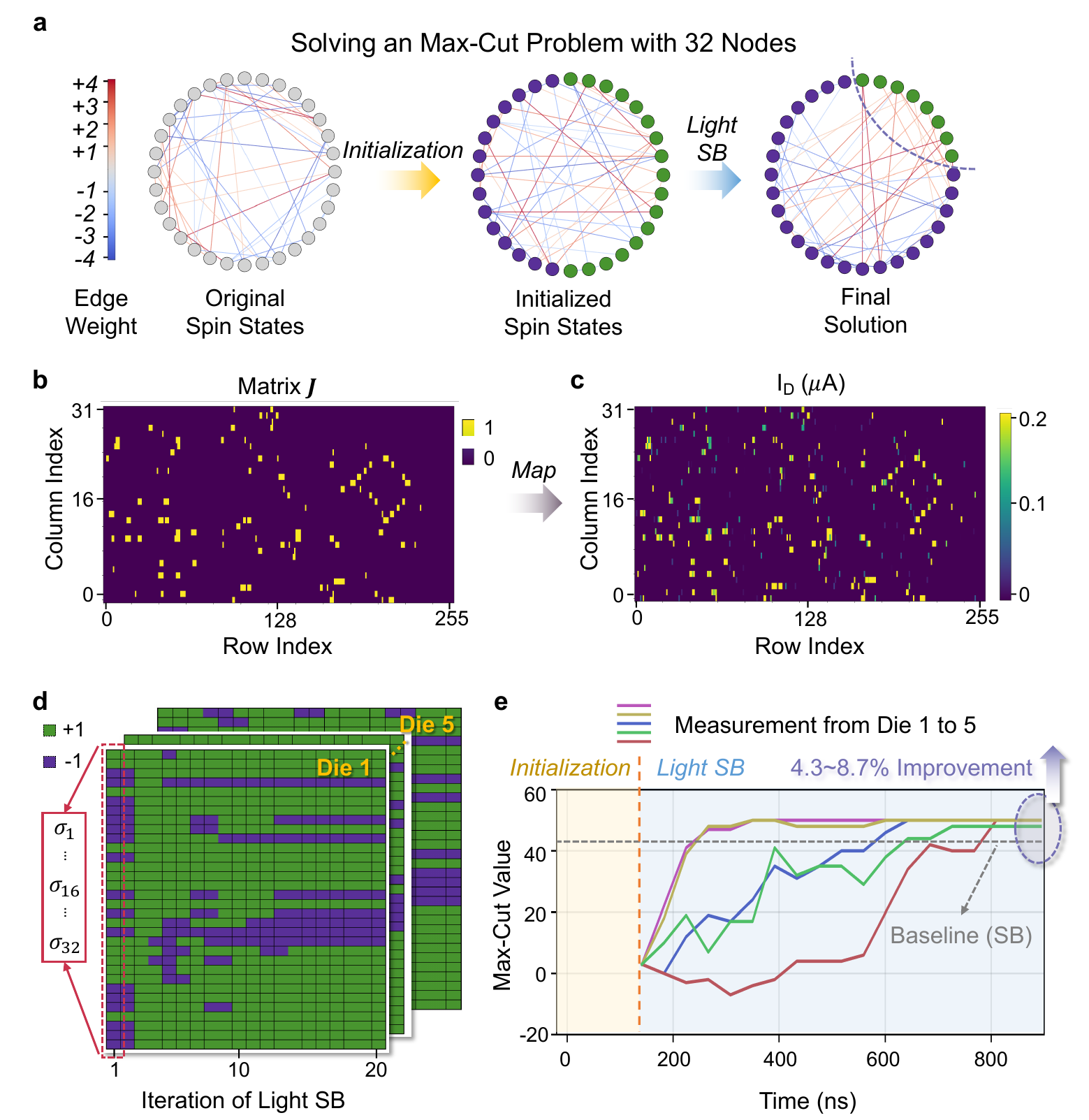}
	\caption{ 
    \stdfig{
    \textbf{Demonstration of solving an Max-Cut problem with 32 nodes.}
    \textbf{a.} A 32-node, 10\% density Max-Cut problem is solved. The value of edges are randomly assigned from $\{-4, -3, -2, -1, 1, 2, 3, 4\}$. 
    This problem is solved through the attention-inspired initialization and light SB algorithm implemented on the proposed CiM chip. 
    \textbf{b.} The matrix $\bm{J}$ of Ising model is mapped as \textit{V}\textsubscript{TH} onto the $32\times 256$ FeFET-based CiM chip, represented as a current map. 
    \textbf{c.} Experiments conducted on five independent dies demonstrate that all achieve solutions 4.3-8.7\% better than SB, with convergence reached within 900ns. 
    \textbf{d.} The spins evolve to an optimal solution after 20 iterations. 
    }
	}
	\label{fig:demo}
\end{figurehere}

\begin{table*}[h]
\caption{Summary of Ising Machines}
\label{table:summary}
\centering
\resizebox{\columnwidth}{!}{
\begin{tabular}{|c|c|c|c|c|c|c|c|}
\hline\hline
Reference&\cite{cai2020power}&\cite{yang2020transiently}&\cite{shin2018hardware}&\cite{mahmoodi2019analog}&\cite{hong2021memory}&\cite{yin2024ferroelectric}&This work\\
\hline
\multirow{3}{*}{Problem}&\multirow{3}{*}{Max-Cut}&\multirow{3}{*}{Max-Cut}&\multirow{2}{*}{Spin}&\multirow{2}{*}{Graph}&\multirow{2}{*}{Traveling}&Graph Coloring/&\multirow{3}{*}{Max-Cut}\\
& & &\multirow{2}{*}{Glass}&\multirow{2}{*}{Partion}&\multirow{2}{*}{Salesman}&Max-Cut/&\\
& & & & & &Prime Factorization&\\
\hline
Initialization&\multirow{2}{*}{Random}&\multirow{2}{*}{Random}&\multirow{2}{*}{Random}&\multirow{2}{*}{Random}&\multirow{2}{*}{Random}&\multirow{2}{*}{Random}&\multirow{2}{*}{Attention-inspired}\\
Strategy& & & & & & & \\
\hline
Method&SA&Chaotic SA&SA&SA&Multi-step SA&MESA&Light SB\\
\hline
Hardware&memristor&memristor&RRAM&RRAM&RRAM&FeFET& BEOL FeFET\\
Implementation&based crossbar&based crossbar&based crossbar&based crossbar&based crossbar&based crossbar&based crossbar\\
\hline
Hardware Size&60$\times$60&2$\times$2&11$\times$3&64$\times$64&1024$\times$1152&32$\times$32&32$\times$256$^\dagger$\\
\hline
Problem Size&60 node&5 node&15 node&6 node&100 node&21 node&32 node\\
\hline
\end{tabular}
}
\begin{flushleft}
\scriptsize
$^\dagger$: The largest FeFET based CiM array known to us.
\end{flushleft}
\end{table*}

\section*{\textcolor{black}{Discussion}}
\label{sec:conclusion}

\std{
We have proposed a comprehensive hardware-algorithm co-design framework for efficient complex COPs solving. 
The solver comprises a two-step algorithmic flow, 
including the attention-inspired initialization followed by a hardware-friendly light SB algorithm, 
which jointly optimizes the solver's ability to converge rapidly and reach superior solutions. 
The fabrication and deployment of the first BEOL $32 \times 256$ FeFET CiM chip for accelerating the core vector-matrix and vector-matrix-vector operation validate the key hardware component of this framework. 
This co-designed Ising machine solver 
achieves up to 175.9$\times$ speedup over a GPU-based conventional SB implementation, while simultaneously  delivering higher-quality solution  (up to 6.54\% improvement in Max-Cut values). 
The summary in Table \ref{table:summary} clearly demonstrates that our rigorously co-designed hardware-software framework 
outperforms representative COP solvers 
reported in prior works. 
Moreover, the proposed algorithmic flow is highly general.
The attention-inspired initialization scheme can be readily integrated with other optimization algorithms, 
such as SA, to enhance their convergence behavior.
Likewise,  the light SB algorithm maintains strong hardware-affinity and can be efficiently deployed on other CiM-based Ising machines and digital computers. 
The experimentally demonstrated 32$\times$256 BEOL FeFET CiM chip thus  showcases the substantial potential of FeFET CiM technology for larger and more complex COP solving. 
}

\section*{Methods}
\subsection*{FeFET Chip integration}
\textcolor{black}{The CiM chip is designed with FeFETs integrated on 180\,nm CMOS platform. The fabricated ferroelectric field effect transistor (FeFET) features a BEOL integrated TiN/Hf\textsubscript{x}Zr\textsubscript{1-x}O\textsubscript{2} (10 nm)/TiN capacitor stacked on top of a front end of line Si transistor. For the Si transistor, a TiN metal gate electrode was deposited using physical vapor deposition (PVD), on top of which the poly-Si gate electrode is deposited. The source and drain n+ regions were obtained by phosphorous ion implantation, which were then activated by a rapid thermal
annealing (RTA) at approximately 1000$^\circ$C. The ferroelectric gate stack process module starts with the growth of a thin TiN\textsubscript{2} based interfacial layer on the TiN capped AlSiCu metal layer, followed by the deposition of a 10\,nm thick doped Hf\textsubscript{x}Zr\textsubscript{1-x}O\textsubscript{2}. The ferroelectric anneal hereby supports the BEOL integration with an anneal thermal budget $<$400°C. }

\subsection*{FeFET Chip Electrical Characterization}
\textcolor{black}{The measurements primarily utilize a PXIe measurement system provided by National Instruments. A padring comprising 30 individual analog and digital pads establishes connections between the 8kb (32$\times$256) FeFET macro and a serial peripheral interface (SPI). The adapter board interfaces with specific pads on the wafer using a probe-card within a wafer probe system. The setup includes a distinct NI PXIe-4163 source measure unit (SMU) and an NI PXIe-6570 pattern generator. Notably, the output pins of the latter device can function as a Pin Parametric Measurement Unit (PPMU). This configuration facilitates the generation of necessary supply, bias voltages, and digital signals. Moreover, the pattern generator plays a crucial role in configuring the scan chain for the proper addressing of wordlines and sourceline/drainline.
}

\bibliography{ref}

@inproceedings{ni2018soc,
  title={SoC logic compatible multi-bit FeMFET weight cell for neuromorphic applications},
  author={Ni, K and Smith, JA and Grisafe, B and Rakshit, T and Obradovic, B and Kittl, JA and Rodder, M and Datta, S},
  booktitle={2018 IEEE International Electron Devices Meeting (IEDM)},
  pages={13--2},
  year={2018},
  organization={IEEE}
}

@article{lucas2014ising,
  title={Ising formulations of many NP problems},
  author={Lucas, Andrew},
  journal={Frontiers in physics},
  volume={2},
  pages={5},
  year={2014},
  publisher={Frontiers}
}

@inproceedings{takemoto20214,
  title={A 144Kb annealing system composed of 9$\times$ 16Kb annealing processor chips with scalable chip-to-chip connections for large-scale combinatorial optimization problems},
  author={Takemoto, Takashi and Yamamoto, Kasho and Yoshimura, Chihiro and Hayashi, Masato and Tada, Masafumi and Saito, Hiroaki and Mashimo, Mayumi and Yamaoka, Masanao},
  booktitle={2021 IEEE International Solid-State Circuits Conference (ISSCC)},
  volume={64},
  pages={64--66},
  year={2021},
  organization={IEEE}
}

@article{dutta2021ising,
  title={An Ising Hamiltonian solver based on coupled stochastic phase-transition nano-oscillators},
  author={Dutta, Suryendy and Khanna, Abhishek and Assoa, AS and Paik, Hanjong and Schlom, Darrell G and Toroczkai, Zolt{\'a}n and Raychowdhury, Arijit and Datta, Suman},
  journal={Nature Electronics},
  volume={4},
  number={7},
  pages={502--512},
  year={2021},
  publisher={Nature Publishing Group UK London}
}

@article{cai2020power,
  title={Power-efficient combinatorial optimization using intrinsic noise in memristor Hopfield neural networks},
  author={Cai, Fuxi and Kumar, Suhas and Van Vaerenbergh, Thomas and Sheng, Xia and Liu, Rui and Li, Can and Liu, Zhan and Foltin, Martin and Yu, Shimeng and Xia, Qiangfei and Yang, Joshua J. and Beausoleil, Raymond and Lu, Wei D. and Strachan, John Paul},
  journal={Nature Electronics},
  volume={3},
  number={7},
  pages={409--418},
  year={2020},
  publisher={Nature Publishing Group}
}

@article{yang2020transiently,
  title={Transiently chaotic simulated annealing based on intrinsic nonlinearity of memristors for efficient solution of optimization problems},
  author={Yang, Ke and Duan, Qingxi and Wang, Yanghao and Zhang, Teng and Yang, Yuchao and Huang, Ru},
  journal={Science advances},
  volume={6},
  number={33},
  pages={eaba9901},
  year={2020},
  publisher={American Association for the Advancement of Science}
}

@inproceedings{shin2018hardware,
  title={Hardware acceleration of simulated annealing of spin glass by RRAM crossbar array},
  author={Shin, Jong Hoon and Jeong, Yeon Joo and Zidan, Mohammed A and Wang, Qiwen and Lu, Wei D},
  booktitle={2018 IEEE International Electron Devices Meeting (IEDM)},
  pages={3--3},
  year={2018},
  organization={IEEE}
}

@inproceedings{mahmoodi2019analog,
  title={An analog neuro-optimizer with adaptable annealing based on 64$\times$ 64 0T1R crossbar circuit},
  author={Mahmoodi, MR and Kim, H and Fahimi, Z and Nili, H and Sedov, Leonid and Polishchuk, Valentin and Strukov, DB},
  booktitle={2019 IEEE International Electron Devices Meeting (IEDM)},
  pages={14--7},
  year={2019},
  organization={IEEE}
}

@inproceedings{hong2021memory,
  title={In-Memory Annealing Unit (IMAU): Energy-Efficient (2000 TOPS/W) Combinatorial Optimizer for Solving Travelling Salesman Problem},
  author={Hong, Ming-Chun and Cho, Le-Chih and Lin, Chih-Sheng and Lin, Yu-Hui and Chen, Po-An and Wang, I-Ting and Tzeng, Pei-Jer and Sheu, Shyh-Shyuan and Lo, Wei-Chung and Wu, Chih-I and Hou, Tuo-Hung},
  booktitle={2021 IEEE International Electron Devices Meeting (IEDM)},
  pages={21--3},
  year={2021},
  organization={IEEE}
}

@book{yu2013industrial,
  title={Industrial applications of combinatorial optimization},
  author={Yu, Gang},
  volume={16},
  year={2013},
  publisher={Springer Science \& Business Media}
}

@book{paschos2014applications,
  title={Applications of combinatorial optimization},
  author={Paschos, Vangelis Th},
  volume={3},
  year={2014},
  publisher={John Wiley \& Sons}
}

@article{naseri2020application,
  title={Application of combinatorial optimization strategies in synthetic biology},
  author={Naseri, Gita and Koffas, Mattheos AG},
  journal={Nature communications},
  volume={11},
  number={1},
  pages={2446},
  year={2020},
  publisher={Nature Publishing Group UK London}
}

@article{barahona1988application,
  title={An application of combinatorial optimization to statistical physics and circuit layout design},
  author={Barahona, Francisco and Gr{\"o}tschel, Martin and J{\"u}nger, Michael and Reinelt, Gerhard},
  journal={Operations Research},
  volume={36},
  number={3},
  pages={493--513},
  year={1988},
  publisher={INFORMS}
}

@article{markov2014limits,
  title={Limits on fundamental limits to computation},
  author={Markov, Igor L},
  journal={Nature},
  volume={512},
  number={7513},
  pages={147--154},
  year={2014},
  publisher={Nature Publishing Group UK London}
}

@article{markov2013know,
  title={Know your limits (review of" limits to parallel computation: p-completeness theory"; greenlaw, r., et al; 1995)[book review]},
  author={Markov, Igor L},
  journal={IEEE Design \& Test},
  volume={30},
  number={1},
  pages={78--83},
  year={2013},
  publisher={IEEE}
}

@book{greenlaw1995limits,
  title={Limits to parallel computation: P-completeness theory},
  author={Greenlaw, Raymond and Hoover, H James and Ruzzo, Walter L},
  year={1995},
  publisher={Oxford University Press on Demand}
}

@article{moy20221,
  title={A 1,968-node coupled ring oscillator circuit for combinatorial optimization problem solving},
  author={Moy, William and Ahmed, Ibrahim and Chiu, Po-wei and Moy, John and Sapatnekar, Sachin S and Kim, Chris H},
  journal={Nature Electronics},
  volume={5},
  number={5},
  pages={310--317},
  year={2022},
  publisher={Nature Publishing Group UK London}
}

@inproceedings{katsuki2022fast,
  title={Fast Solving Complete 2000-Node Optimization Using Stochastic-Computing Simulated Annealing},
  author={Katsuki, Kota and Shin, Duckgyu and Onizawa, Naoya and Hanyu, Takahiro},
  booktitle={2022 29th IEEE International Conference on Electronics, Circuits and Systems (ICECS)},
  pages={1--4},
  year={2022},
  organization={IEEE}
}

@article{pierangeli2019large,
  title={Large-scale photonic Ising machine by spatial light modulation},
  author={Pierangeli, D and Marcucci, G and Conti, C},
  journal={Physical review letters},
  volume={122},
  number={21},
  pages={213902},
  year={2019},
  publisher={APS}
}

@inproceedings{roychowdhury2021bistable,
  title={Bistable latch Ising machines},
  author={Roychowdhury, Jaijeet},
  booktitle={Unconventional Computation and Natural Computation: 19th International Conference, UCNC 2021, Espoo, Finland, October 18--22, 2021, Proceedings 19},
  pages={131--148},
  year={2021},
  organization={Springer}
}

@misc{maxcutstanford,
  title = {{Stanford Max-Cut dataset}},
  note = {\url{https://web.stanford.edu/~yyye/yyye/Gset/}}
}

@article{onizawa2023local,
  title={Local Energy Distribution Based Hyperparameter Determination for Stochastic Simulated Annealing},
  author={Onizawa, Naoya and Kuroki, Kyo and Shin, Duckgyu and Hanyu, Takahiro},
  journal={arXiv preprint arXiv:2304.11839},
  year={2023}
}

@article{inagaki2016coherent,
  title={A coherent Ising machine for 2000-node optimization problems},
  author={Inagaki, Takahiro and Haribara, Yoshitaka and Igarashi, Koji and Sonobe, Tomohiro and Tamate, Shuhei and Honjo, Toshimori and Marandi, Alireza and McMahon, Peter L and Umeki, Takeshi and Enbutsu, Koji and others},
  journal={Science},
  volume={354},
  number={6312},
  pages={603--606},
  year={2016},
  publisher={American Association for the Advancement of Science}
}

@inproceedings{afoakwa2021brim,
  title={BRIM: bistable resistively-coupled Ising machine},
  author={Afoakwa, Richard and Zhang, Yiqiao and Vengalam, Uday Kumar Reddy and Ignjatovic, Zeljko and Huang, Michael},
  booktitle={2021 IEEE International Symposium on High-Performance Computer Architecture (HPCA)},
  pages={749--760},
  year={2021},
  organization={IEEE}
}

@article{honjo2021100,
  title={100,000-spin coherent Ising machine},
  author={Honjo, Toshimori and Sonobe, Tomohiro and Inaba, Kensuke and Inagaki, Takahiro and Ikuta, Takuya and Yamada, Yasuhiro and Kazama, Takushi and Enbutsu, Koji and Umeki, Takeshi and Kasahara, Ryoichi and others},
  journal={Science advances},
  volume={7},
  number={40},
  pages={eabh0952},
  year={2021},
  publisher={American Association for the Advancement of Science}
}

@article{yamamoto2020coherent,
  title={Coherent Ising machines—Quantum optics and neural network Perspectives},
  author={Yamamoto, Y and Leleu, T and Ganguli, S and Mabuchi, H},
  journal={Applied Physics Letters},
  volume={117},
  number={16},
  pages={160501},
  year={2020},
  publisher={AIP Publishing LLC}
}

@article{mcmahon2016fully,
  title={A fully programmable 100-spin coherent Ising machine with all-to-all connections},
  author={McMahon, Peter L and Marandi, Alireza and Haribara, Yoshitaka and Hamerly, Ryan and Langrock, Carsten and Tamate, Shuhei and Inagaki, Takahiro and Takesue, Hiroki and Utsunomiya, Shoko and Aihara, Kazuyuki and Byer, Robert L. and Fejer, M. M. and Mabuchi, Hideo and Yamamoto, Yoshihisa},
  journal={Science},
  volume={354},
  number={6312},
  pages={614--617},
  year={2016},
  publisher={American Association for the Advancement of Science}
}

@article{bohm2019poor,
  title={A poor man’s coherent Ising machine based on opto-electronic feedback systems for solving optimization problems},
  author={B{\"o}hm, Fabian and Verschaffelt, Guy and Van der Sande, Guy},
  journal={Nature communications},
  volume={10},
  number={1},
  pages={3538},
  year={2019},
  publisher={Nature Publishing Group UK London}
}

@article{ahmed2021probabilistic,
  title={A probabilistic compute fabric based on coupled ring oscillators for solving combinatorial optimization problems},
  author={Ahmed, Ibrahim and Chiu, Po-Wei and Moy, William and Kim, Chris H},
  journal={IEEE Journal of Solid-State Circuits},
  volume={56},
  number={9},
  pages={2870--2880},
  year={2021},
  publisher={IEEE}
}

@article{mallick2023cmos,
  title={CMOS-compatible ising machines built using bistable latches coupled through ferroelectric transistor arrays},
  author={Mallick, Antik and Zhao, Zijian and Bashar, Mohammad Khairul and Alam, Shamiul and Islam, Md Mazharul and Xiao, Yi and Xu, Yixin and Aziz, Ahmedullah and Narayanan, Vijaykrishnan and Ni, Kai and others},
  journal={Scientific reports},
  volume={13},
  number={1},
  pages={1515},
  year={2023},
  publisher={Nature Publishing Group UK London}
}

@article{tatsumura2021scaling,
  title={Scaling out Ising machines using a multi-chip architecture for simulated bifurcation},
  author={Tatsumura, Kosuke and Yamasaki, Masaya and Goto, Hayato},
  journal={Nature Electronics},
  volume={4},
  number={3},
  pages={208--217},
  year={2021},
  publisher={Nature Publishing Group UK London}
}

@inproceedings{yamamoto20207,
  title={STATICA: A 512-spin 0.25 M-weight full-digital annealing processor with a near-memory all-spin-updates-at-once architecture for combinatorial optimization with complete spin-spin interactions},
  author={Yamamoto, Kasho and Ando, Kota and Mertig, Normann and Takemoto, Takashi and Yamaoka, Masanao and Teramoto, Hiroshi and Sakai, Akira and Takamaeda-Yamazaki, Shinya and Motomura, Masato},
  booktitle={2020 IEEE International Solid-State Circuits Conference-(ISSCC)},
  pages={138--140},
  year={2020},
  organization={IEEE}
}

@article{cilasun2025coupled,
  title={A coupled-oscillator-based Ising chip for combinatorial optimization},
  author={C{\i}lasun, H{\"u}srev and Moy, William and Zeng, Ziqing and Islam, Tahmida and Lo, Hao and Vanasse, Alex and Tan, Megan and Anees, Mohammad and Kumar, Abhimanyu and Sapatnekar, Sachin S and others},
  journal={Nature Electronics},
  pages={1--10},
  year={2025},
  publisher={Nature Publishing Group}
}

@article{yue2024scalable,
  title={A scalable universal Ising machine based on interaction-centric storage and compute-in-memory},
  author={Yue, Wenshuo and Zhang, Teng and Jing, Zhaokun and Wu, Kai and Yang, Yuxiang and Yang, Zhen and Wu, Yongqin and Bu, Weihai and Zheng, Kai and Kang, Jin and others},
  journal={Nature Electronics},
  volume={7},
  number={10},
  pages={904--913},
  year={2024},
  publisher={Nature Publishing Group UK London}
}

@article{qian2025ferroelectric,
  title={Ferroelectric Compute-in-Memory Framework for Solving Pure and Mixed Strategy Nash Equilibrium},
  author={Qian, Yu and Huang, Ding and Vardar, Alptekin and Laleni, Nellie and Zhou, Min and Ni, Kai and K{\"a}mpfe, Thomas and Zhuo, Cheng and Yin, Xunzhao},
  journal={IEEE Transactions on Circuits and Systems I: Regular Papers},
  year={2025},
  publisher={IEEE}
}

@inproceedings{qian2024c,
  title={C-Nash: A Novel Ferroelectric Computing-in-Memory Architecture for Solving Mixed Strategy Nash Equilibrium},
  author={Qian, Yu and Ni, Kai and Kampfe, Thomas and Zhuo, Cheng and Yin, Xunzhao},
  booktitle={Proceedings of the 61st ACM/IEEE Design Automation Conference},
  pages={1--6},
  year={2024}
}

@inproceedings{qian2024hycim,
  title={HyCiM: A Hybrid Computing-in-Memory QUBO Solver for General Combinatorial Optimization Problems with Inequality Constraints},
  author={Qian, Yu and Yang, Zeyu and Ni, Kai and Vardar, Alptekin and Kampfe, Thomas and Yin, Xunzhao},
  booktitle={Proceedings of the 61st ACM/IEEE Design Automation Conference},
  pages={1--6},
  year={2024}
}

@article{yin2024ferroelectric,
  title={Ferroelectric compute-in-memory annealer for combinatorial optimization problems},
  author={Yin, Xunzhao and Qian, Yu and Vardar, Alptekin and G{\"u}nther, Marcel and M{\"u}ller, Franz and Laleni, Nellie and Zhao, Zijian and Jiang, Zhouhang and Shi, Zhiguo and Shi, Yiyu and others},
  journal={Nature Communications},
  volume={15},
  number={1},
  pages={2419},
  year={2024},
  publisher={Nature Publishing Group UK London}
}

@article{qian2025device,
  title={Device-Algorithm Co-Design of Ferroelectric Compute-in-Memory In-Situ Annealer for Combinatorial Optimization Problems},
  author={Qian, Yu and Huang, Xianmin and Wang, Ranran and Yang, Zeyu and Zhou, Min and K{\"a}mpfe, Thomas and Zhuo, Cheng and Yin, Xunzhao},
  journal={arXiv preprint arXiv:2504.21280},
  year={2025}
}

@article{dee2024design,
  title={Design of a Mixed-Signal Compute-in-Memory Ising Solver With Sub-$\mu$s Time-to-Solution and Optimal Decaying Noise Profile},
  author={Dee, Alana Marie and Vuong, Duy and Bennett, Katherine and Moazeni, Sajjad},
  journal={IEEE Transactions on Circuits and Systems I: Regular Papers},
  year={2024},
  publisher={IEEE}
}

@article{jiang2023efficient,
  title={Efficient combinatorial optimization by quantum-inspired parallel annealing in analogue memristor crossbar},
  author={Jiang, Mingrui and Shan, Keyi and He, Chengping and Li, Can},
  journal={Nature communications},
  volume={14},
  number={1},
  pages={5927},
  year={2023},
  publisher={Nature Publishing Group UK London}
}

@article{goto2019combinatorial,
  title={Combinatorial optimization by simulating adiabatic bifurcations in nonlinear Hamiltonian systems},
  author={Goto, Hayato and Tatsumura, Kosuke and Dixon, Alexander R},
  journal={Science advances},
  volume={5},
  number={4},
  pages={eaav2372},
  year={2019},
  publisher={American Association for the Advancement of Science}
}

@article{goto2021high,
  title={High-performance combinatorial optimization based on classical mechanics},
  author={Goto, Hayato and Endo, Kotaro and Suzuki, Masaru and Sakai, Yoshisato and Kanao, Taro and Hamakawa, Yohei and Hidaka, Ryo and Yamasaki, Masaya and Tatsumura, Kosuke},
  journal={Science Advances},
  volume={7},
  number={6},
  pages={eabe7953},
  year={2021},
  publisher={American Association for the Advancement of Science}
}

@inproceedings{zhang2024high,
  title={A high-performance stochastic simulated bifurcation Ising machine},
  author={Zhang, Tingting and Zhang, Hongqiao and Yu, Zhengkun and Liu, Siting and Han, Jie},
  booktitle={Proceedings of the 61st ACM/IEEE Design Automation Conference},
  pages={1--6},
  year={2024}
}

@article{hassanat2018improved,
  title={An improved genetic algorithm with a new initialization mechanism based on regression techniques},
  author={Hassanat, Ahmad B and Prasath, VB Surya and Abbadi, Mohammed Ali and Abu-Qdari, Salam Amer and Faris, Hossam},
  journal={Information},
  volume={9},
  number={7},
  pages={167},
  year={2018},
  publisher={MDPI}
}

@article{vaswani2017attention,
  title={Attention is all you need},
  author={Vaswani, Ashish and Shazeer, Noam and Parmar, Niki and Uszkoreit, Jakob and Jones, Llion and Gomez, Aidan N and Kaiser, {\L}ukasz and Polosukhin, Illia},
  journal={Advances in neural information processing systems},
  volume={30},
  year={2017}
}

@misc{yset,
  title = {{Generated Large-Scale Max-Cut Dataset}},
  note = {\url{https://web.stanford.edu/~yyye/yyye/Gset/}}
}

\bibliographystyle{Nature}

\newpage
\renewcommand{\thefigure}{S\arabic{figure}}
\renewcommand{\thetable}{S\arabic{table}}

\onecolumn
\centering
\textbf{\Large Supplementary Information}
\setcounter{figure}{0}
\setcounter{table}{0}
\setcounter{page}{1}

\begin{flushleft} 
\section{\textbf{\large Simulated Bifurcation}}
\label{sec:light SB}
\subsection{Conventional Simulated Bifurcation}
\label{subsec: SB}
\end{flushleft}

\justify 
\std{
The simulated bifurcation (SB) algorithm, proposed in \cite{goto2019combinatorial, goto2021high}, aims to minimize the Ising energy, i.e., $\min E = \bm{\sigma^T} J \bm{\sigma}$. 
Here, $J$ presents the coupling matrix derived from the problem, and $\bm{\sigma}$ is the input variable vector, where $\sigma_i \in\{1, -1\}$ represents the state of spin $i$. 
In SB, each spin $i$ is represented by two continuous variables: its position $\bm{X}_i \in [-1,1]$ and momenta $\bm{Y}_i\in [-1,1]$. These two vectors, i.e., $\bm{X}$ and $\bm{Y}$, are updated iteratively based on each other according to the following equations: 
\begin{equation}
\label{equ: supp SB Y}
    \bm{Y} = \bm{Y} - K\bm{X}^3 - (\Delta - p)\bm{X} + \zeta \bm{J}\bm{X}
\end{equation}
\begin{equation}
\label{equ: supp SB X}
    \bm{X} = \bm{X} + \Delta \bm{Y}
\end{equation}
}
\std{
Here, $K$, $\Delta$ and $\zeta$ are constants, with values aligning with the  settings in \cite{goto2019combinatorial}. 
The parameter $p$ is continuously increased from $0$ to $\Delta$ iteratively. 
}

\subsection{Impact of Quantization Interval}
\label{subsec: interval}

\begin{figurehere}
	\centering
	\includegraphics [width=0.8\linewidth]{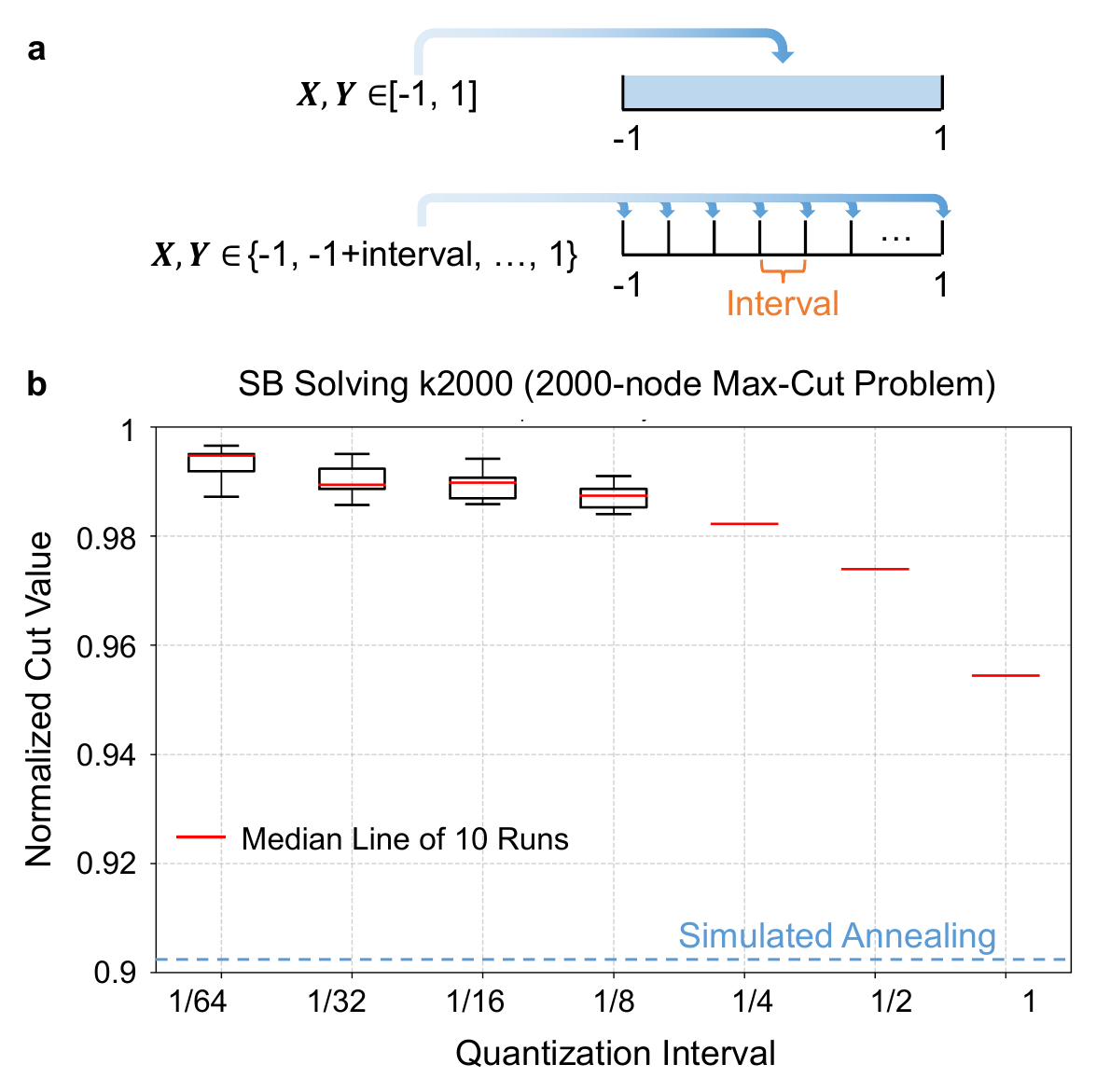}
 \caption{
 \std{
 \textbf{Impact of quantization in SB.}
 \textbf{a.} Quantization process of $\bm{X}$ and $\bm{Y}$. 
 \textbf{b.} In solving a 2000-node Max-Cut problem with different quantization intervals, the average cut-value of SB is decreased as the interval increases.
 }}
\label{fig: interval}
\end{figurehere}

\justify 
\std{
The conventional SB algorithm requires floating-point quantization for its position and momentum vectors, i.e., $\bm{X}$ and $\bm{Y}$ $\in [-1,1]$, which poses a significant burden on hardware implementation. 
To address this, we investigated the impact of lower-bit quantization on solution quality (cut value).  
As shown in Fig. \ref{fig: interval}\textbf{a}, the continuous interval of $\bm{X}$ and $\bm{Y}$ is narrowed to a discrete set of values, i.e., $\{-1, -1+\text{interval}, ..., 1\}$, where "interval" is a configurable parameter. 
We then evaluated the solution quality of SB on the K2000 Max-Cut problem \cite{inagaki2016coherent}, a 2000-node graph. For this benchmark, we performed 10 independent runs for each of seven different interval sizes: $1/64, 1/32, 1/16, 1/8, 1/4, 1/2, 1$. 
Fig. \ref{fig: interval}\textbf{b} shows the normalized cut values under each quantization configuration. 
The results are normalized against the floating-point quantization result, while the simulated annealing (SA) result serves as a baseline. 
As the interval size increases, the normalized cut value shows a slight decrease, ranging from 0.995 for an interval of $1/64$, to 0.954 for an interval of $1$. 
Importantly, all tested quantization configurations yield better results than SA, which achieves a normalized cut value of 0.904. 
To maximize hardware compatibility, we adopted an interval of $1$ in this work. 
An attention-inspired initialization method is proposed 
to compensate for the solution quality loss incurred by this low-bit quantization. 
}

\begin{flushleft} 
\section{\textbf{\large FeFET CiM chip integration \& testing}}
\end{flushleft}

\subsection{Wafer-Level Measurement Setup for CiM Chip}
\label{subsec: measurement}

\begin{figurehere}
	\centering
	\includegraphics [width=0.9\linewidth]{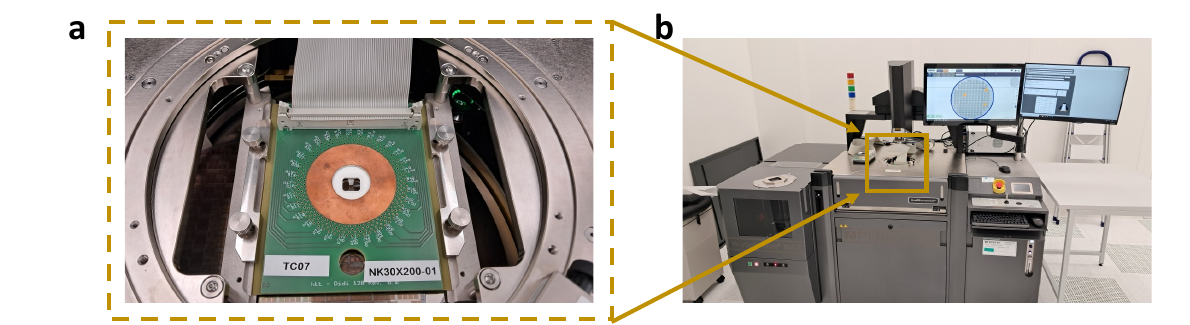}
 \caption{
 \textbf{Wafer-level experimental setup for FeFET characterization.}
 \textbf{a.} Needle card employed to contact the CiM chip pads. 
 \textbf{b.} Wafer prober system MPI TS3500 allowing for the precise contacting and characterization.
 }
\label{fig: chip measure}
\end{figurehere}

\justify 

The electrical characterization setup for wafer-level characterization is shown in Fig. \ref{fig: chip measure}. The CiM chip padring of 30 individual analog and digital pads connect the 8kb (32$\times$256) FeFET macro with a serial peripheral interface (SPI). 
The FeFET testchip (given in Fig. \ref{fig: wafer-level setup}) was tested on wafer-level using an MPI TS3500 with a wafer loader. A set of separate NI PXIe-4163 source measure unit (SMU) and an NI PXIe-6570 pattern generator are employed. Notably, the output pins of the latter device can be utilized as a Pin Parametric Measurement Unit (PPMU). Through this arrangement, the requisite supply, bias voltages, and digital signals are generated. The needle card contact is controlled with a MPI SENTIO control for the customized needle card (see Fig. S3b). The needle card connects via a flatband cable to an adapter board. The CiM chip was then characterized using an NI PXIe using pattern generator for the digital SPI interface of the chip and SMUs for the analog testing voltages.
Using the decoder structures of the CiM chip it is possible to select every single FeFET cell as well as switch off the read-out circuitry to measure the transistor characteristics of every single bitcell.

\begin{figurehere}
	\centering
	\includegraphics [width=0.9\linewidth]{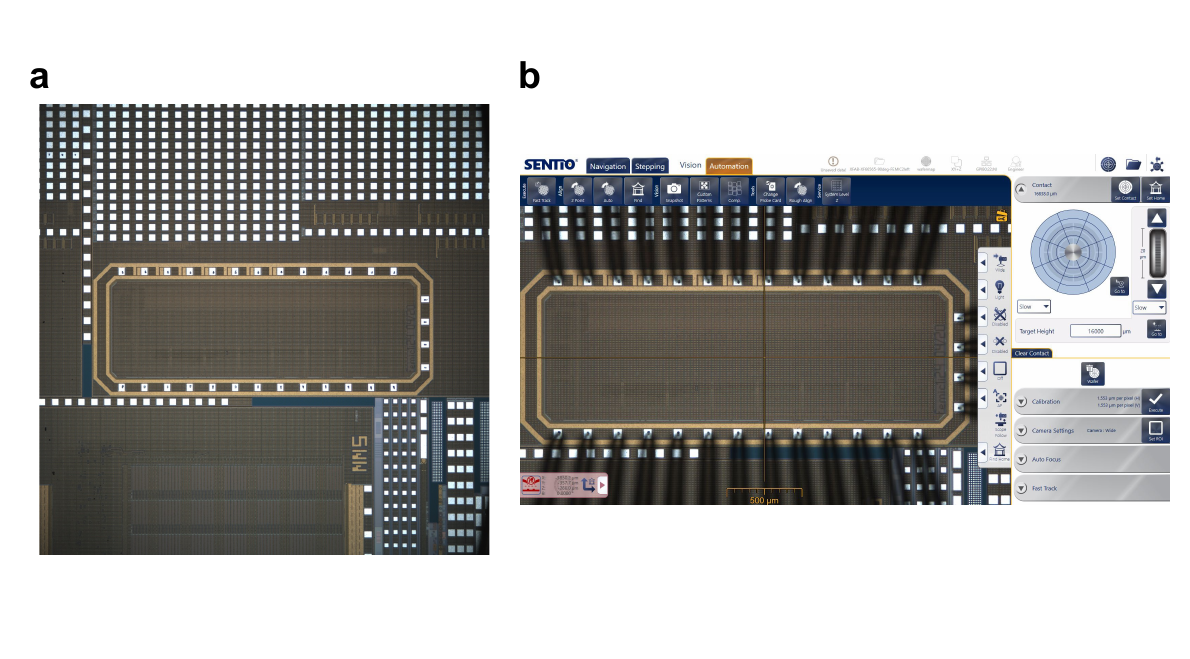}
 \caption{
  \textbf{Wafer-level measurement for CiM chip.}
 \textbf{a.} CiM chip image in wafer probing setup. 
 \textbf{b.} Needle card contacted CiM chip.
 }
\label{fig: wafer-level setup}
\end{figurehere}

\begin{flushleft} 
\section{\textbf{\large Implementation of Ising Machine}}
\subsection{Hardware Setup}
The architecture of the Ising machine including the CiM chip (FEMIC) was implemented on board, supporting its analog and digital SPI interface. The CiM chip was wire-bonded onto a carrier board. The current readout and FeFET programming is performed using the on-board analog-digital converters (ADC) as well as digital-analog-converters (DAC). To support the required 5V voltage regime of the FeFET, level shifters where implemented. The system is controlled with a ESP32 microcontroller.
\label{subsec: system setup}
\end{flushleft}

\begin{figurehere}
	\centering
	\includegraphics [width=0.9\linewidth]{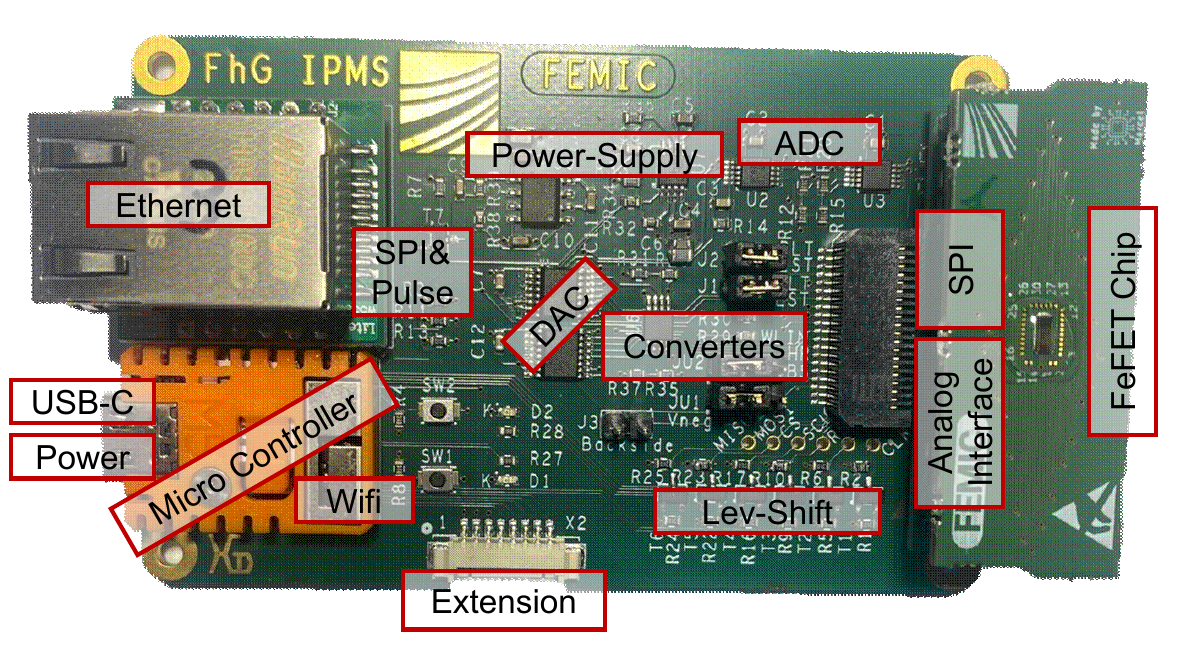}
 \caption{
 \yu{
 \textbf{Board-level details of the proposed Ising machine.}
 }}
\label{fig: board}
\end{figurehere}

\justify

\end{document}